\documentclass{ws-p8-50x6-00}
\begin{document}

\title{
 Quasiparticle description of deconfined matter
 at finite \boldmath$\mu$ and \boldmath$T$}

\author{A.~Peshier, B.~K\"ampfer}

\address{
 Research Center Rossendorf Inc., PF 510119, 01314 Dresden, Germany}

\maketitle

\abstracts{
 An effective quasiparticle description of deconfined QCD thermodynamics
 compatible with both finite temperature nonperturbative lattice data and
 the asymptotic limit is generalized to finite chemical potential.
 Implications for the $N_f = 4$ flavor lattice data extended to $\mu \neq 0$
 as well as for deconfined matter with realistic quark masses are considered.
}

 Matter in or close to local thermodynamic equilibrium can be described by
 its equation of state (EoS), to study, e.\,g., the cosmologic evolution or
 the dynamics of heavy-ion collisions in the framework of hydrorelativistic
 equations.
 In the examples mentioned, for strong excitations, the expected transition
 of hadronic matter to a plasma state of deconfined quarks and gluons, and
 hence, its EoS are of central importance.
 The coupling in the physically relevant regime being large, nonperturbative
 methods as lattice QCD are required to describe deconfined matter reliably.
 As a matter of fact, these simulations are presently restricted to vanishing
 chemical potential $\mu$, and the implementation of physical quark masses
 is still too expensive numerically, so the EoS is not yet known from first
 principles.
 Here we generalize a quasiparticle approach, which reproduces finite
 temperature lattice results, and present realistic extensions of EoS's
 to $\mu \neq 0$.

 Asymptotically, the collective behavior of the plasma can be interpreted
 in terms of quasiparticle excitations with a dispersion relation
 $\omega_i^2(k) \approx m_i^2+k^2$ and $m_i^2=m_{0i}^2+\Pi_i^*$
 depending on the rest mass $m_{0i}$ and the leading order on-shell
 selfenergies\cite{LeBellac},
 \bea
  \Pi_q^*
  &=&
  2\, \omega_{q0}\, (m_0 + \omega_{q0}) \, ,
  \quad
  \omega_{q0}^2
  =
  \frac{N_c^2-1}{16N_c}\, \left[ T^2+\frac{\mu_q^2}{\pi^2} \right] G^2 \, ,
  \nonumber \\
  \Pi_g^*
  &=&
  \frac16
  \left[
     \left( N_c+\frac12\, N_f \right) T^2
   + \frac{N_c}{2\pi^2} \sum_q \mu_q^2
  \right] G^2 \, .
 \label{Pi}
 \eea
 Decomposed into the various quasiparticle contributions and their
 interaction $B$, the pressure of the quasiparticle system
 \be
  p(T,\mu; m_{0i}^2)
  =
  \sum_j p_j^{\rm id}(T, \mu_j(\mu); m_j^2) - B(\Pi_i^*)
  \label{p_qp}
 \ee
 is stationary with respect to the selfenergies\cite{GY}, $\partial p /
 \partial \Pi_i^* = \partial p_i^{\rm id} / \partial m_i^2 - \partial B /
 \partial \Pi_i^* = 0$, and represents an effective resummation of the
 leading-order thermal contributions\cite{TFT}.
 By comparison to lattice data, at $\mu=0$, this intuitive picture has
 been shown\cite{PR_LH} to hold even close to confinement, with the
 effective coupling in (\ref{Pi}) nonperturbatively parametrized by
 \be
   G^2(T,\mu=0)
   =
   48\pi^2
   \left[
    (11 N_c - 2\, N_f)
     \ln\left(\frac{T+T_s}{T_c/\lambda}\right)^{\!\!2}\,
   \right]^{-1} ,
   \label{G2}
 \ee
 interpolating to the asymptotic limit of QCD.

 In the following, the approach is generalized to finite values of $\mu$,
 where, as a consequence of the stationarity of the pressure, $B$ is
 determined in differential form by
 \be
  \frac{\partial B}{\partial T}
  =
  \sum_j
   \frac{\partial p^{\rm id}}{\partial m_j^2}\,
   \frac{\partial \Pi_j^*}{\partial T}
  \equiv
  B_T
  \, ,  \qquad
  \frac{\partial B}{\partial \mu}
  =
  \sum_j
   \frac{\partial p^{\rm id}}{\partial m_j^2}\,
   \frac{\partial \Pi_j^*}{\partial \mu}
  \equiv
  B_\mu \, .
  \label{B_x}
 \ee
 The functions $B_T$ and $B_\mu$ have to respect the integrability condition
 \be
   \frac{\partial B_T}{\partial \mu} - \frac{\partial B_\mu}{\partial T}
  =
  \sum_j
  \left[
    \frac{\partial n_j^{\rm id}}{\partial m_j^2}\,
     \frac{\partial \Pi_j^*}{\partial T}
   -\frac{\partial s_j^{\rm id}}{\partial m_j^2}\,
     \frac{\partial \Pi_j^*}{\partial \mu}
  \right]
  =
  0 \, ,
  \label{flow}
 \ee
 which features a flow equation for $G^2(T,\mu)$.
 Together with (\ref{Pi},\ref{p_qp}) it allows to extend the EoS from
 $\mu=0$, calculated, e.\,g., by lattice QCD and parametrized by
 $G^2(T,0)$, to finite values of $\mu$ in a thermodynamically consistent
 way.

 The approach is first applied to the model case of a QCD plasma with
 $N_f=4$ light quark flavors. At vanishing chemical potential, the EoS
 calculated on the lattice\cite{EoS_4f} in a restricted interval of $T$
 is reproduced by the effective coupling (\ref{G2}) with $\lambda=6.7$
 and $T_s/T_c=-0.81$ as well as appropriate numbers of quasiparticle
 degrees of freedom.
 In the weak coupling limit, $m_i^2/T^2 \sim G^2 \rightarrow g^2
 \rightarrow 0$, the equation (\ref{flow}) reduces to
 \be
  \pi^2 \left( c T^2 + \frac{\mu^2}{\pi^2} \right)
  \frac1\mu\, \frac{\partial g^2}{\partial \mu}
  -
  \left( T^2 + \frac{\mu^2}{\pi^2} \right)
  \frac1T\, \frac{\partial g^2}{\partial T}
  =
  0 \, ,
  \quad
  c = \frac{4 N_c + 5 N_f}{9 N_f} \, ,
 \ee
 which yields an elliptic flow with $g^2$ being constant along the
 characteristics $c T^4 + 2 T^2 (\mu/\pi)^2 + (\mu/\pi)^4 = const.$
 In the nonperturbative regime shown in Figure 1, the pronounced
 increase of $G^2$ towards the region of small $T,\mu$ reflects the
 vicinity of confinement, while asymptotically the coupling decreases
 as expected.
 It is worth to note that the pressure $p(T,\mu)$ of the system
 becomes negative for certain values of $\mu$ at small temperatures.
 This region of instability provides a lower boundary for the value
 of the confining chemical potential $\mu_c \ge 3.4\, T_c$ at $T=0$.

 Addressing a physically relevant question, the EoS of deconfined quark
 gluon matter with vanishing net strangeness, $\mu_s=0$, is estimated in
 the quasiparticle approach for $\mu_u=\mu_d=\mu$, with realistic quark
 masses taken into account.
 Until precise lattice data may be available for $\mu=0$, the unknown
 effective coupling $G^2(T,\mu=0)$ can be parametrized by reasonable
 choices of $\lambda$ and $T_s/T_c$, which are constrained by the Gibbs
 condition at the finite temperature confinement transition $T_c \approx
 150\,$MeV.
 The thermodynamic potential $p(T,\mu)$ with a representative choice of
 the model parameters is shown in Figure 2. The confinement region is
 characterized by a distinct reduction of the effective degrees of freedom.
 As known qualitatively from lattice results at $\mu=0$, the energy density
 $e=T\,\partial p / \partial T + \mu\,\partial p / \partial \mu - p$
 approaches some 80...90\% of its asymptotic value already near the
 confinement region, leading there to a considerable softening of the EoS.
 Due to the stationarity of the pressure, $\partial p / \partial \Pi_i^*$ = 0,
 these features are to a large extent insensitive to the model parameters
 so the approach provides realistic estimates for the EoS of deconfined
 matter with detailed consequences for heavy-ion collisions to be studied
 in hydrorelativistic models.

\begin{figure}[hbt]
\begin{minipage}{5.7cm}
  \epsfxsize 5.7cm
  \centerline{\epsffile{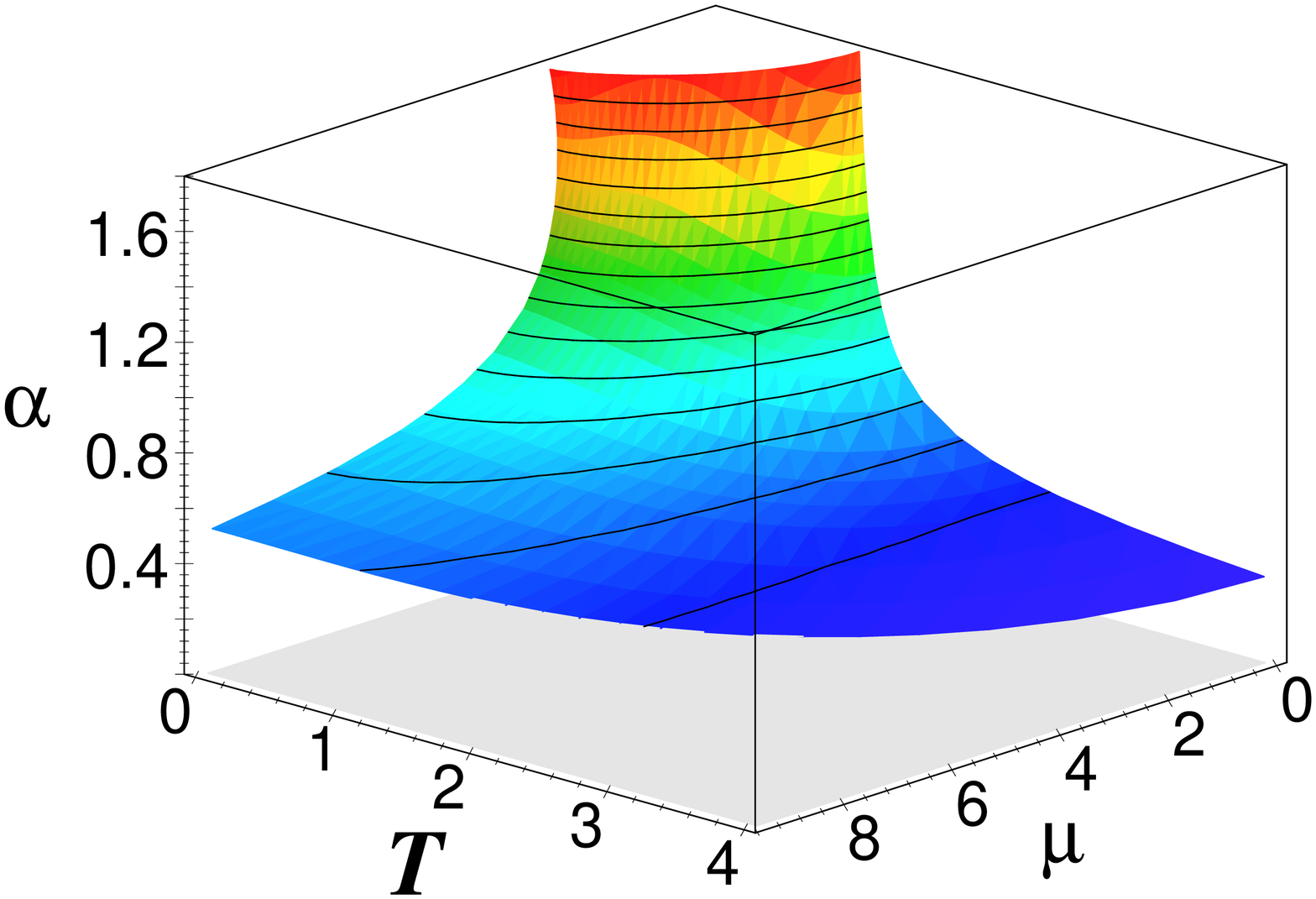}}
  \caption{The effective coupling $\alpha=G^2/4\pi$ for the
          $N_f=4$ QCD plasma.}
\end{minipage}
\hfill
\begin{minipage}{5.7cm}
  \epsfxsize 5.7cm
  \centerline{\epsffile{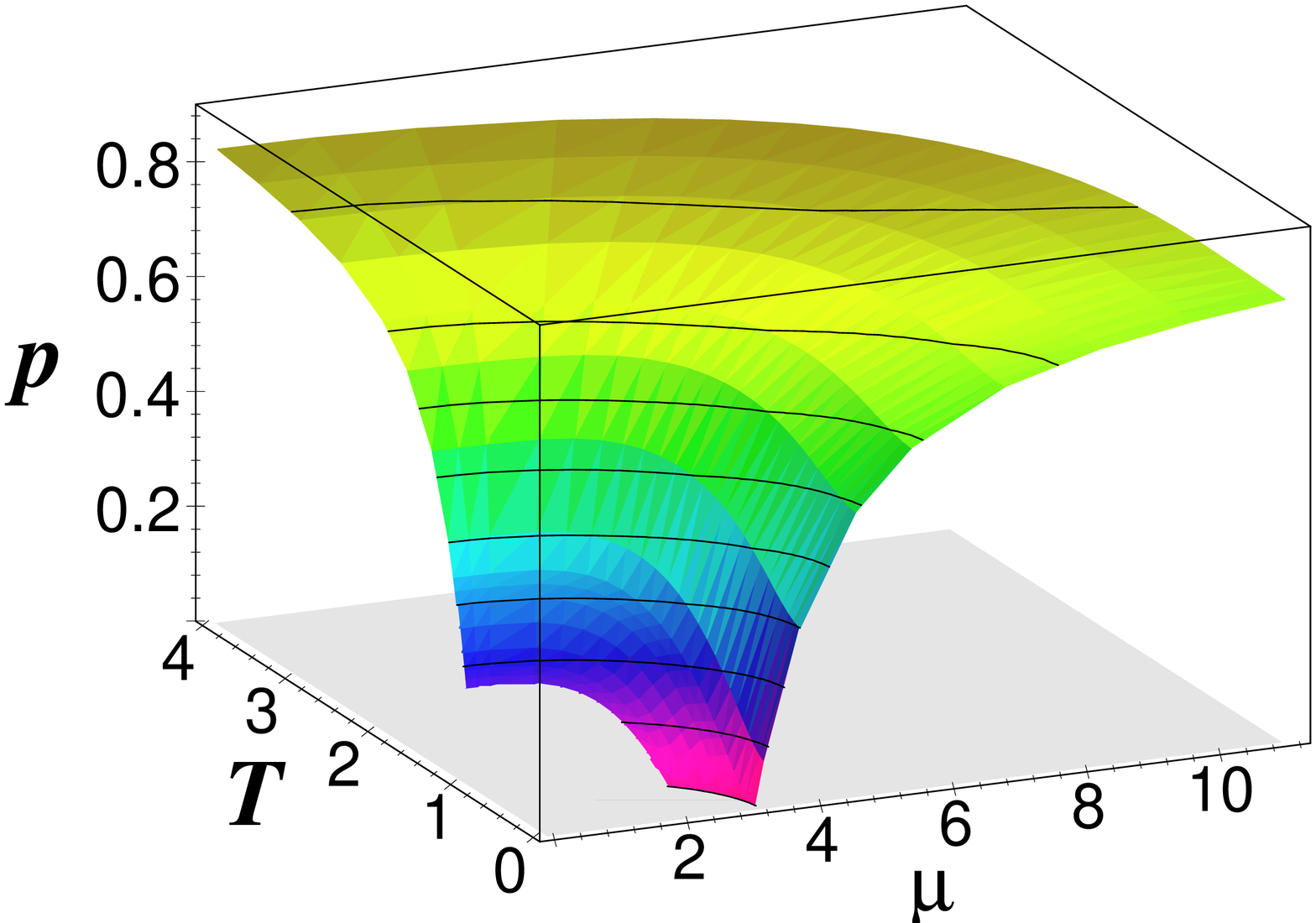}}
  \caption{The pressure of the (2+1) quark gluon plasma scaled by the asymptotic limit.}
\end{minipage}
\end{figure}


\begin{thebibliography}{99}
\bibitem{LeBellac}
   M.~Le Bellac, {\em Thermal Field Theory},
   Cambridge University Press (1996)
\bibitem{GY}
   M.\,I.~Gorenstein, S.\,N.~Yang,
   Phys.\ Rev.\ D52 (1995) 5206
\bibitem{TFT}
   A.~Peshier,
   TFT'98 Proceedings, hep-ph/9809379
\bibitem{PR_LH}
   A.~Peshier, B.~K\"ampfer, O.\,P.~Pavlenko, G.~Soff,
   Phys.\ Rev.\ D54 (1996) 2399,
   P.~Levai, U.~Heinz,
   Phys.\ Rev.\ C57 (1998) 1879
\bibitem{EoS_4f}
   J.~Engels, R.~Joswig, F.~Karsch, E.~Laermann, M.~L\"utgemeier,
   B.~Petersson,
   Phys.\ Lett.\ B396 (1997) 210
\end{thebibliography}
\end{document}